\documentclass[journal]{IEEEtran}

\usepackage{amsmath}
\usepackage{amsthm}
\usepackage{amssymb}
\usepackage{amsfonts}
\usepackage{enumitem}
\usepackage{graphicx}
\usepackage{url}
\usepackage{subfigure}
\usepackage{float}
\usepackage{mathrsfs,fourier,pifont}
\usepackage{threeparttable}
\usepackage{algorithmic}
\usepackage{algorithm}
\usepackage{color}
\usepackage{booktabs}
\usepackage{epstopdf}
\usepackage{multicol,multienum, multirow}
\usepackage{makecell}
\usepackage[numbers,sort&compress]{natbib}
\usepackage{threeparttable}
\usepackage[utf8]{inputenc}
\newsavebox{\tablebox}

\begin{document}

\title{The Effects of Using Taxi-Hailing Application on Driving Performance}

\author{Xiexing~Feng,
	Libo~Cao,
	Yunxian~Zhang,
	Hongbo~Gao,
    Lifan~Tan

	\thanks{X. Feng is with the State Key Laboratory of Advanced Design and Manufacturing for Vehicle Body, Hunan University, Changsha, Hunan, China, and also with the Department of Electrical and Computer Engineering, University of Windsor, Windsor, ON N9B 3P4, Canada.}
	\thanks{L. Cao, Y. Zhang and L. Tan are with the State Key Laboratory of Advanced Design and Manufacturing for Vehicle Body, Hunan University, Changsha, Hunan, China.}
	\thanks{H. Gao is wiht the State Key Laboratory of Automotive Safety and Energy, Tsinghua University, Beijing, China}
	\thanks{L. Cao and H. Gao are the corresponding authors. (e-mail: hdclb@163.com; ghb48@mail.tsinghua.edu.cn)}
	
}

\maketitle

\begin{abstract}
Driver distraction has become a major threat to the road safety, and the globally booming taxi-hailing application introduces new source of distraction to drivers. Although various in-vehicle information systems (IVIS) have been studied extensively, no documentation exists objectively measuring the extent to which interacting with taxi-hailing application during driving impacts drivers' behavior. To fill this gap, a simulator-based study was conducted to synthetically compare the effects that different output modalities (visual, audio, combined visual-audio) and input modalities (baseline, manual, speech) imposed on the driving performance. The results show that the visual output introduced more negative effects on driving performance compared to audio output. In the combined output, visual component dominated the effects imposed on the longitudinal control and hazard detection; audio component only exacerbated the negative effects of visual component on the lateral control. Speech input modality was overall less detrimental to driving performance than manual input modality, especially reflected in the drivers' quicker reaction to hazard events. The visual-manual interaction modality most severely impaired the hazard detecting ability, while also led to strong compensative behaviors. The audio-speech and visual-speech modality associated with more smooth lateral control and faster response to hazard events respectively compared to other modality. These results could be applied to improve the design of not only the taxi-hailing application, but also other input-output balanced IVIS.
\end{abstract}

\section{Introduction}
Distraction and inattention of the drivers have become an increasing threat to the road safety. The National Highway Traffic Safety Administration (NATSA) reported that nearly 50\% of crashes concerned with distracted driving. An even closer relationship between distracted driving and road accident was reported by the Naturalistic 100-car Study, revealing that 78\% of crashes and 65\% of near-crashes involved some form of driver inattention \cite{klauer_impact_2006}. Among various forms of inattention including secondary tasks related or driving related inattention, non-specific eye glances, and fatigue \cite{liang_combining_2010}, the proportion of distraction associated with in-vehicle information systems was increasing greatly because their soaring prevalence and rapid innovation. In recent years, the newly invented vehicle sharing application emerged and spread worldwide. It was reported that two types of vehicle sharing application – taxi-hailing application and instant car-booking application respectively had a user scale of up to 159 million and 122 million in China. Although interacting with such applications are driving-related for commercial drivers, among whom there are both professional and ordinary drivers, these interactions may interfere with safe driving and should be still considered distracting \cite{lee_defining_2009}. Furthermore, the interaction mechanism of taxi-hailing application is slightly more complicated than that of the other one, and thus more like to impair the driving performance. So, the taxi-hailing application is worth studying and discussing in detail in order to diminish its potential risk to safety driving.

\begin{table*}[!htb]
	\small\sf\centering
	\caption{The GOMS description of the interactions with four different in-vehicle information systems.}
	\begin{tabular}{l}
		\toprule
		The GOMS description of IVIS \\
		\midrule
		Goal: handle a taxi-hailing order (when accepting the order) \\
		\hspace{8pt} Goal: get the order information \\
		\hspace{16pt} Move the eyes to smart phone screen (V) or listen to broadcast of order information (A) \\
		\hspace{16pt} Memorize the departure and destination (C) \\
		
		\hspace{8pt} Goal: thinking and deciding whether or not to take the order (C)\\
		
		\hspace{8pt} Goal: operate in the application \\
		\hspace{16pt} Lift the right arm (M) \\
		\hspace{16pt} Point to the ‘accept’ button (V, M) \\
		\hspace{16pt} Touch the screen (M) \\
		
		\hline
		
		Goal: turn on the iPod and select a specific song in the song titles menu \cite{chisholm_effects_2008} \\
		\hspace{8pt} Goal: turn on the ipod \\
		\hspace{16pt} Lift the right arm (M) \\
		\hspace{16pt} Move the eyes to the ipod (V) \\
		\hspace{16pt} Point to the power button (V, M) \\
		\hspace{16pt} Press the button (M) \\
		
		\hspace{8pt} Goal: select a specific song \\
		\hspace{16pt} Enter the song titles menu (V, M) \\
		\hspace{16pt} Recognize the song titles (V) \\
		\hspace{16pt} Slides the menu if the song is not found (M) \\
		\hspace{16pt} Match the song title (C) \\
		\hspace{16pt} Press the play button (M) \\
		
		\hline
		
		Goal: reply a text message \cite{hosking_effects_2009} \\
		\hspace{8pt} Goal: get the content of the text message \\
		\hspace{16pt} Move the eyes to phone screen (V) \\
		\hspace{16pt} Memorize the message content (C) \\
		
		\hspace{8pt} Goal: think about problem in the message to get the answer (C) \\
		
		\hspace{8pt} Goal: edit and send the message \\
		\hspace{16pt} Type the answer using the keyboard of phone (V, M, C) \\
		\hspace{16pt} Press the send button (V, M) \\
		
		\hline
		Goal: use navigation system as driving route guidance  \cite{jensen_studying_2010} \\
		\hspace{8pt} Goal: input the destination into the navigation system (before the driving start) \\
		\hspace{16pt} Type the destination using the touch screen (V, M, C) \\
		\hspace{16pt} Touch enter button (V, M) \\
		
		\hspace{8pt} Goal: get the guidance information \\
		\hspace{16pt} Move the eyes to the screen (V) \\
		\hspace{16pt} Recognize the direction and distance (C) \\
		
		\bottomrule	
	\end{tabular}
	\begin{tablenotes}
	\item Notes: V represents visual resources, A represents auditory resources, C represents cognitive resources, and M represents manual response.
	\end{tablenotes}
	\label{table 1}
\end{table*}

A substantial literature has examined the effects of the interactions with different IVIS on driving performance. Conversation over phone could result in a reduction in headway time, poor lane maintenance, and delayed reaction to driving events \cite{hancock_distraction_2003,irwin_effect_2000,lamble_cognitive_1999,papadakaki_driving_2016,reed_comparison_1999}. \cite{hosking_effects_2009} found that text messaging increased drivers' variability in lane position and the mean time headway up to 50\%, and the headway variability up to 150\%. \cite{papadakaki_driving_2016} acknowledged the increase of variability in lane position, while reported a decrease of the average following distance. \cite{chisholm_effects_2008} reported that difficult iPod interactions, which involved finding a specific song within the song titles menu, significantly increased the perception response time to driving hazards. However, there is no documentation existing that objectively measures the extent to which interacting with taxi-hailing application impacts driving performance and safety. Although some similar characteristics could be identified, drivers' interaction pattern with taxi-hailing application differs from that with other IVIS. The demand of resources by various IVIS could be precisely defined by the GOMS description, as shown in Table 1. The interaction with taxi-hailing application requires balanced demands of input and output, with a wide variety of resources (visual, auditory, cognitive, and manual), but a relatively short occupation time. The interaction with GPS navigation is highly output-oriented towards the driver and less input-oriented \cite{jensen_studying_2010}, while that with MP3 is highly input-oriented. Texting message requires both input and output, but its input, namely typing a message, occupies the drivers' resources for a much longer time compared to the input of taxi-hailing application – touching 'accept' button. Therefore, results of previous studies on other IVIS should not be directly applied to the emerging taxi-hailing application. It is essential to conduct a study that specially focuses on the taxi-hailing application in order to understand its influence on drivers' behaviors and then to improve its system design.

As stated in Table 1, taxi drivers diverted their attentions from driving task to taxi-hailing orders in three phases - getting order information, making the decision, and operating the application. Considering the inevitability and diversity of decision making process of different drivers, this paper focuses on how the different information access modes and operation modes, namely the output and input modalities of the taxi-hailing application, affect the driving performance.

Previous researches have studied the different output modalities of some IVIS, especially the GPS navigation guide. \cite{green_examination_1993} conducted a driving simulator study comparing four output modalities of the GPS navigation guide. Results showed that although the auditory output required less attention and gave the lower reaction time when compared to the visual output, participants generally expressed a slight inclination towards visual output. \cite{jensen_studying_2010} evaluated how three different output configurations – audio, visual, and audio-visual – of a GPS system affected driving behavior. Results illustrated that visual output not only caused a substantial amount of off-road glances, but also impaired drivers' longitudinal and lateral control. When supplementing the visual configurations with audio output, the number of off-road glances decreased but driving performance was not affected. Besides the researches on the output techniques of in-vehicle systems, there were also a mass of literatures that focused on the input modalities. \cite{maciej_comparison_2009} conducted a study to determine whether speech-based interfaces for IVIS reduce the distraction caused by these systems. Results revealed that although speech interfaces improved driving performance for most systems, these improvements were overall not strong enough to reach the baseline performance level. A study of \cite{tornros_mobile_2005} focused on effects of dialing and conversation using handsfree and handheld mobile phone in simulated driving. Different phone modes affected the lateral position deviation in the same pattern; in contrast, only the handheld phone mode caused a decrease in speed, which could be interpreted as a compensatory effort for the increased mental workload. The vast majority of prior studies either only focused on one aspect of the interaction (input or output), or analyzed both the input and output modalities separately \cite{gellatly_use_1997,hosking_effects_2009,tsimhoni_address_2004}; however, little efforts were made to determine the integrated effects of the combinations of different input and output modalities.

The goal of the current study is to synthetically compare the effects that the different combinations of output modalities (visual, audio, combined visual-audio) and input modalities (baseline, manual, speech) of the taxi-hailing application impose the on the driving performance. Based on prior studies, it is hypothesized that audio output introduces less interference with driving than visual output does. Speech input results in less distraction compared to manual input. The visual-manual and audio-speech combinations are expected to respectively impose the most and least impairment on driving performance.

\begin{figure*}
	\centering
	\setlength\tabcolsep{1.5pt} 
	\begin{tabular}{c} 
		\includegraphics[width=0.8\textwidth]{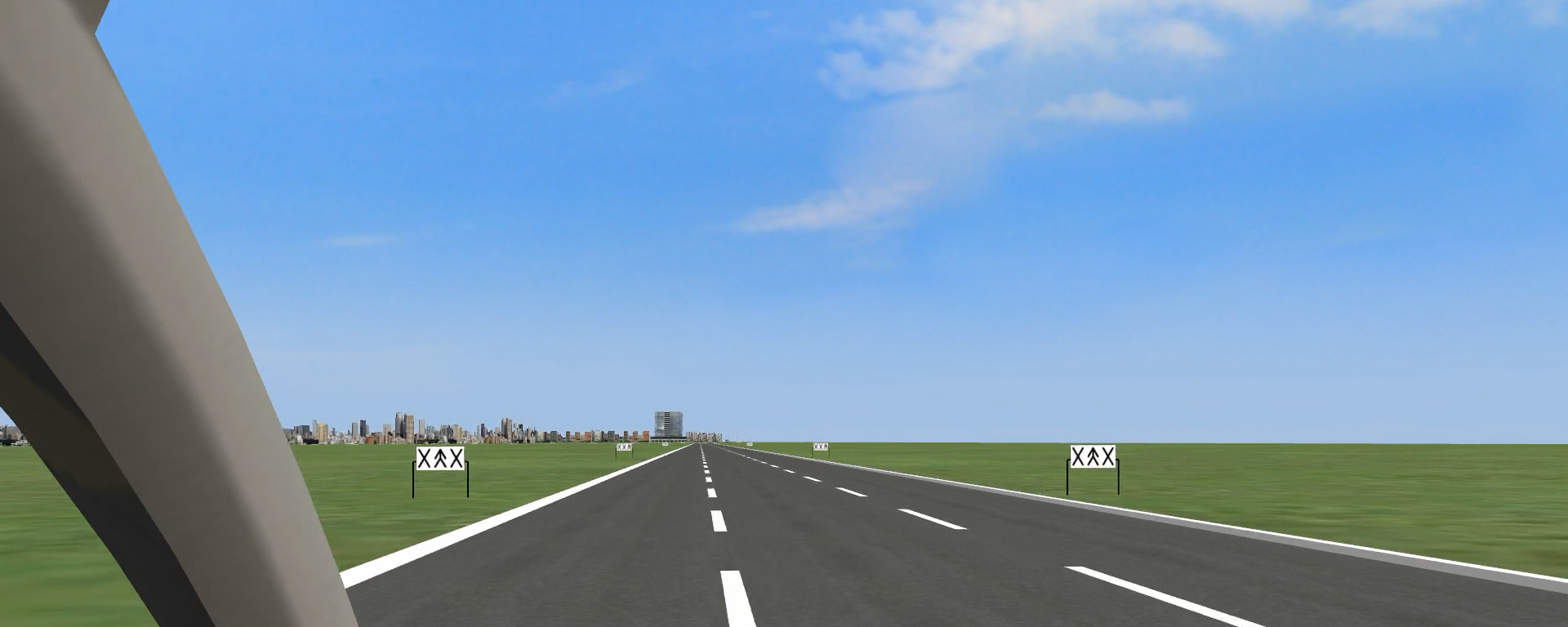}\\
		(a) Lane changing task scene \\
		\includegraphics[width=0.8\textwidth]{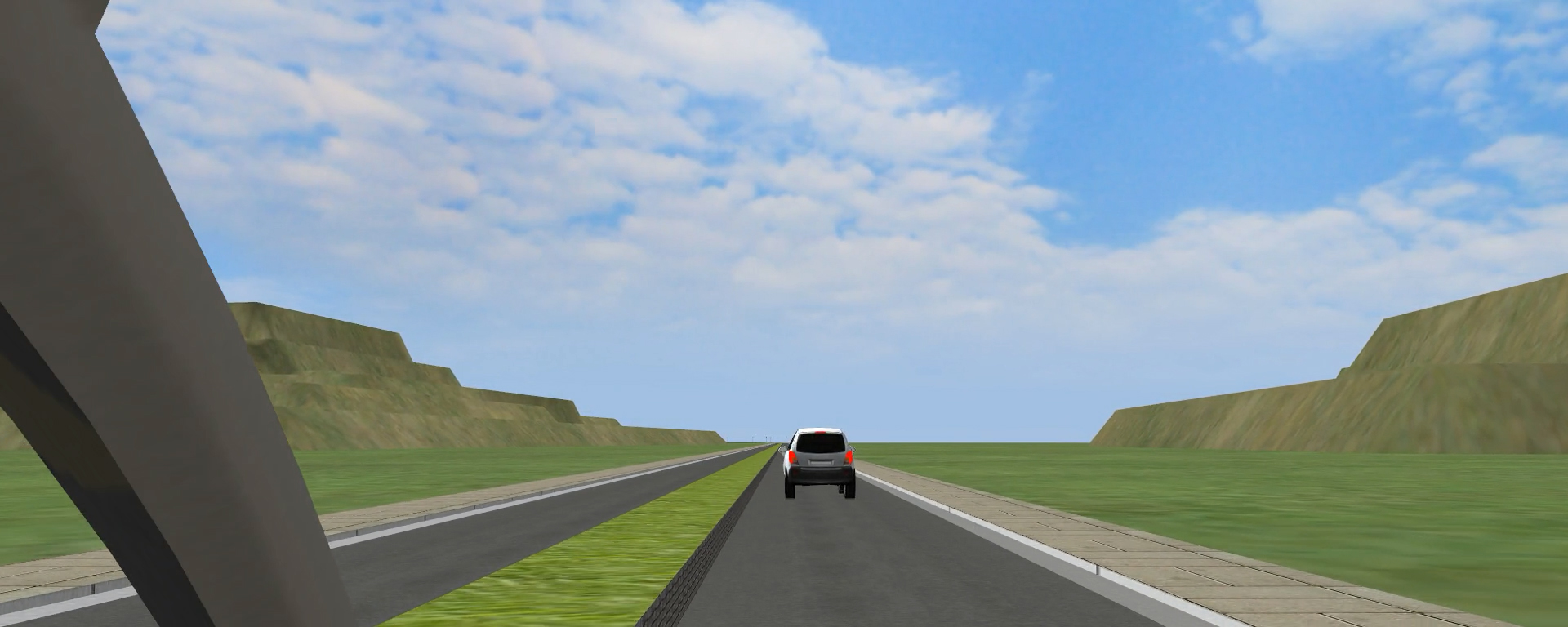}\\
		(b) Car following task scene \\
		\includegraphics[width=0.8\textwidth]{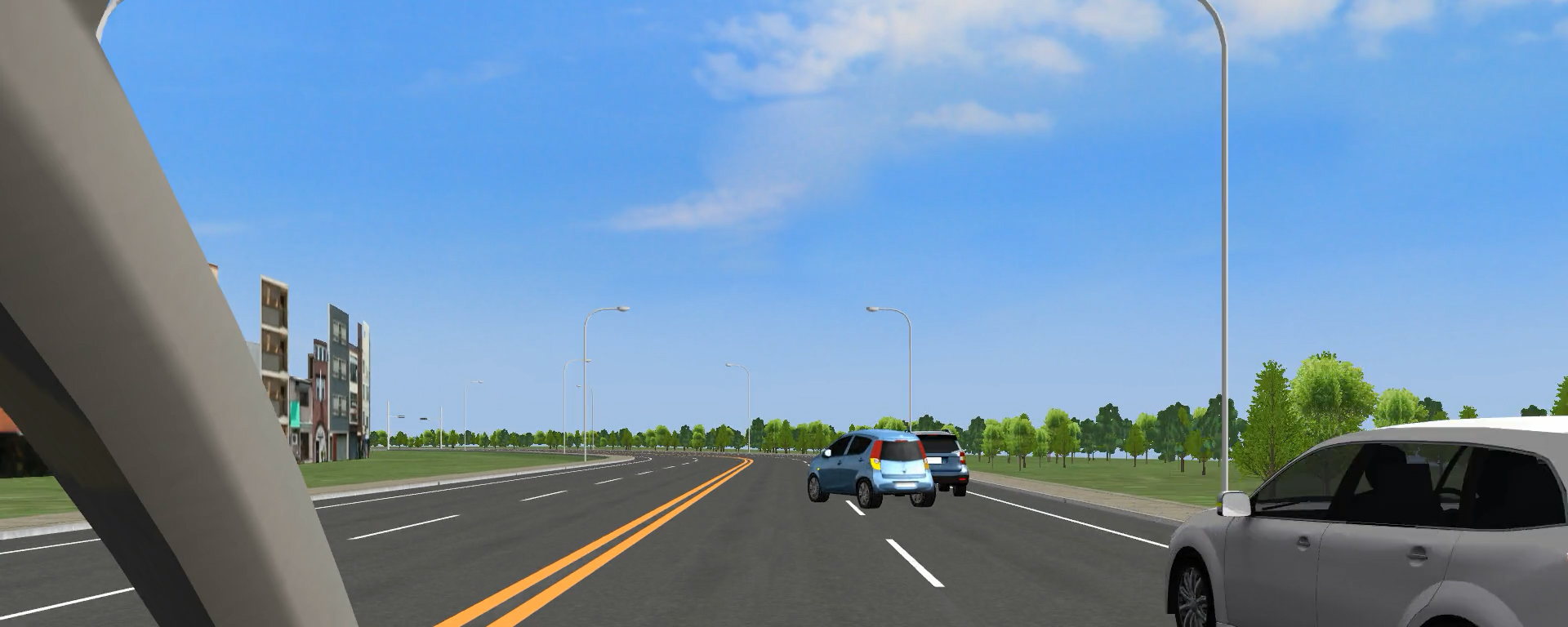}\\
		(c) Hazard event scene \\
	\end{tabular}
	\caption{Virtual driving environment and driving tasks scenes}
\end{figure*}

\begin{figure*}
	\centering
	\setlength\tabcolsep{1.5pt} 
	\begin{tabular}{c} 
		\includegraphics[width=0.8\textwidth]{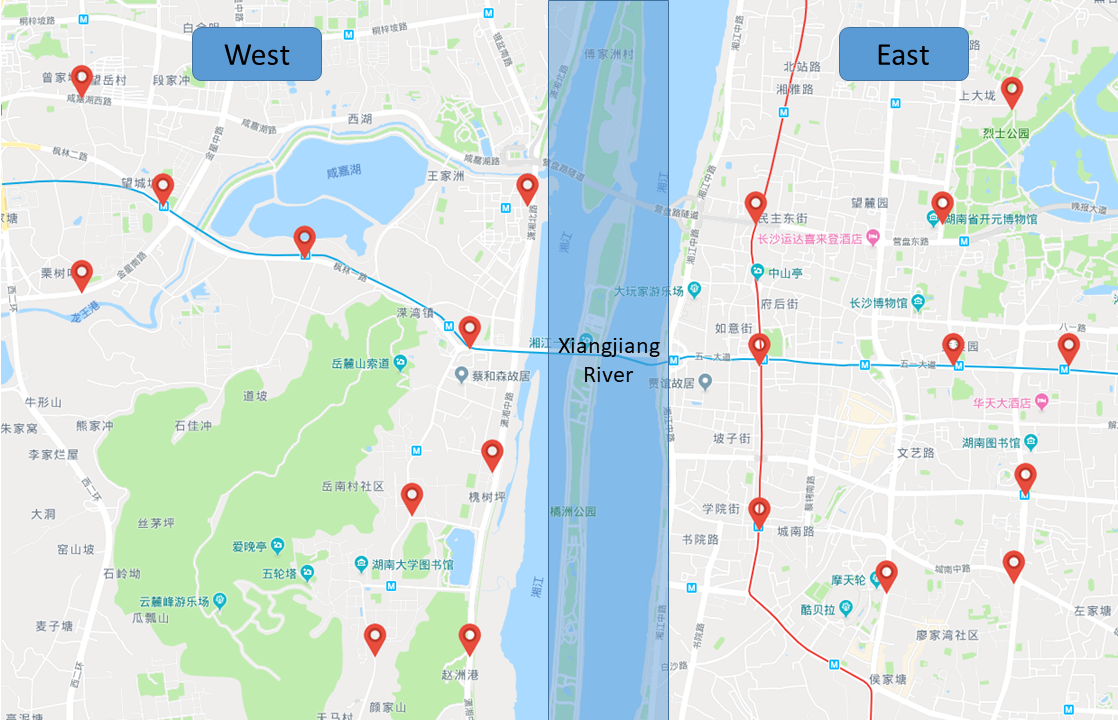}\\
	\end{tabular}
	\caption{Selected locations for the rules of taxi-hailing order decision making}
\end{figure*}

\section{Method}

\subsection{Participants}
Risks of distraction-related crashes have been found to be especially high for young drivers because of their inexperience and the strong willingness to engage in new technology \cite{lee_technology_2007}. So 18 qualified male participants aged between 23 and 27 (M=25.54) and holding a license more than 3 years (M=3.52) were selected. All participants have prior experience with taxi-hailing app and were familiar with its work mechanism.

\subsection{Apparatus}
The experiment took place in a portable driving simulator at Hunan University. Three
42 in. LED screens with $1920\times 1080$ resolution located approximately 0.8m in front of the driver, providing him/her a 120$^\circ$ field of view. Simulator data were sampled at 30Hz. A 5.1 in. smart phone was mounted on the right side of the dash, approximately 30$^\circ$ laterally and 15$^\circ$ vertically below the drivers' line of sight. To provide a controllable and realistic experimental interface, a special taxi-hailing application, which could generate orders according to the experimental requirements, was developed using APP Inventor.

\subsection{Experimental Design}
The experiment was a 3 (input modalities) $\times$ 3 (output modalities) within-subject design. Three input modalities included no operation (baseline), manual modality and speech-based modality, which were distributed in three experimental drives respectively. In the baseline drive, there was no taxi-hailing order and participants were not supposed to conduct any operation other than driving. In the manual and speech drives, participants were instructed to interact with the application by either touching the screen or speaking out. Participants completed each of three drives in turn, which were counterbalanced using a balanced Latin square design. Three levels of output modalities were visual, audio and combined visual-audio. Orders of each output modality were generated 4 times in each drive, and the sequence of them were randomly arranged in each drive. Because there was no order in the baseline drive, the data collecting region of each output modality matched the one at which orders of corresponding modality otherwise would have taken place in manual and speech drives.

\subsection{Driving Environment and Secondary Tasks}
The driving environment was an 8 km circuit, mainly including three parts - lane changing task region, car following task region and normal urban environment, as shown in Figure 1. The lane changing task was a revised version \cite{petzoldt_how_2014} of the standard Lane Change Test (Mattes, 2003), consisted of a 1800m section of a straight three-lane road with 12 lane-changing signs placed on both side of the road at intervals of about 150m (M=150, SD=46.4). Participants were instructed to drive at 60km/h and to switch into the signed lane as soon as they could recognize the sign. In manual and speech drives, orders of three different output modalities occurred when subject vehicle came to 40m before three (among the twelve) lane change signs. In the car following task, participants drove on a straight single-lane road and assigned to maintain a 1.5-s gap from a leading vehicle. The leading vehicle traveled at 50km/h, and transiently slowed down to 20km/h then speeded up back five times. In manual and speech drives, orders took place during three (out of the five) decelerate-acceleration sessions. The normal urban environment mainly consisted of dual-lane straight and curve roads. Participants were instructed to drive as normally as possible according to Chinese road rules and to adhere to 60km/h speed limit. There were nine cars parked on the side of the roadway at different locations. Three cars therein would suddenly pull out with taxi-hailing orders accompanied when the subject vehicle was 58.8m away from them; another one car pulled out without order accompanied and the rest five cars remained stationary, which aimed to reduce participants' anticipation of the occurrence of the hazard events. Besides the nine orders occurred during task session, another three took placed during normal driving. Thus, a total of 12 orders were triggered respectively in manual and speech drives.

A beep signaled to participants that a taxi-hailing order was occurred. For the order of visual-audio modality, the departure and destination were displayed on the screen as well as broadcasted automatically; for visual modality, order information was displayed without broadcast; for audio modality, a piece of pre-recorded naturalistic audio saying the order information was played while no information was displayed on screen.

Participants had 10s to consider and decide whether or not to accept the order; if they decided to take, they should touch the screen or speak out within this 10s, otherwise the order was considered passed. In real-world situation, such decision involved a variety of factors, such as the traveling distance and the traffic situation. However, considering the limited simulation condition, a simplified and controllable rule was specified. 20 locations in Changsha were selected to serve as the departure and destination, with a half in the west to Xiangjiang River and a half in the east, as shown Figure 2. If the departure and destination of an order were on the same side, the order should be taken, and vice versa. This rule simplified the thinking process of judging an order, so the selected locations were with long addresses (about 8 Chinese characters) and seldom used in daily life, in order to raise drivers' mental workload to approximate the realistic situation. In each drive, all of nine orders in the task session were supposed to be accepted because the influence of the operations to take orders (manual or speech) upon driving performance was to be recorded and analyzed; while the rest three orders during normal driving session were randomly arranged.

\subsection{Procedure}
Participants initially read and signed the consent document, and the personal information was reported including age and driving experience. Participants were then given a map showing the 20 locations and the take-or-pass rule, and tested by researcher after 10-min. of remembering. After that, participants practiced operating the simulator and the taxi-hailing application respectively until they felt comfortable, based on self-report. Then participants completed a whole practice drive, and had to accomplish all the driving tasks and to make correct decision for 90\% orders to be qualified. Following the practice drive, participants conducted three experimental drives, each approximately 8-min. long. The whole experimental session was about 75-min. for each participant. The field situation of the experiment is shown in Figure 3.

\subsection{Dependent Variables}
The dependent variables were specified for each event. In hazard events, brake reaction time, approaching speed and minimum headway time were selected. Brake reaction time (BrakeRT) was the time from the starting of the parked car to participants pressing brake pedal. Approaching speed was defined as the average speed of subject vehicle before a hazard event. Minimum headway time (MinHWT) was the minimum value of the distance from the subject vehicle (SV) to the leading vehicle (LV) divided by the velocity of the SV; and a smaller value implied a reduced safety margin and riskier driving behavior. With respect to car following tasks, mean, minimum and variability of headway time were analyzed. For the lane changing events, drivers' vehicle lateral control was characterized by the steering error and steering velocity, and the longitudinal control was characterized by average speed and variability of speed. Steering error was calculated as the absolute difference between the second-order Taylor series expansion prediction of steering angle and the observed steering angle \cite{nakayama_development_1999}, representing the smoothness of steering wheel control. A 15s trial time segment was identified for each task session, and all dependent variables were measured across such segment.

\begin{figure*}
	\centering
	\setlength\tabcolsep{1.5pt} 
	\begin{tabular}{c} 
		\includegraphics[width=0.7\textwidth]{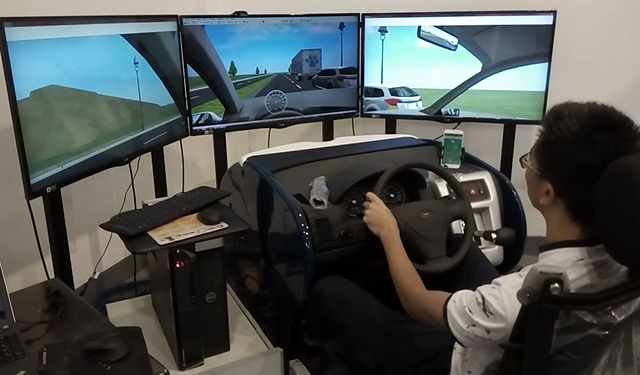}\\
	\end{tabular}
	\caption{The field situation of the experiment}
\end{figure*}

\begin{table*}[htbp]\footnotesize
	\centering
	\caption{ANOVA results of dependent variables of driving performance.}
	\begin{tabular}{lllllllllll}
		\toprule
		Event&	Dependent variables& \multicolumn{9}{l}{Statistical tests}\\
		\cline{3-11}							
		&	&	\multicolumn{3}{l}{Input modality} &		\multicolumn{3}{l}{Output modality}	&		\multicolumn{3}{l}{Interaction} \\
		\cline{3-11}		
		&	&	F&	p&	$\eta^2$&	F&	p&	$\eta^2$&	F&	p&	$\eta^2$\\
		\midrule
		Hazards &	BrakeRT&	6.411&	0.008&	0.274&	0.598&	0.551&	0.034&	4.518&	0.007&	0.210\\
		& Approach Speed&	1.443&	0.251&	0.078&	4.901&	0.021&	0.224&	3.937&	0.02&	0.188\\
		& MinHWT&	2.746&	0.098&	0.139&	0.249&	0.780&	0.014&	5.54&	0.002&	0.246\\
		Car Following& MeanHWT&	4.936&	0.021&	0.225&	2.529&	0.118&	0.130&	0.754&	0.501&	0.042\\
		& MinHWT&	2.041&	0.150&	0.107&	9.067&	0.001&	0.348&	1.962&	0.129&	0.103\\
		& Std. HWT&	2.201&	0.139&	0.115&	2.088&	0.157&	0.109&	0.532&	0.610&	0.030\\
		Lane Changing & 	Steering Error&	1.526&	0.232&	0.082&	18.501&	\textless0.001&	0.521&	2.484&	0.052&	0.127\\
		& Steering Velocity&	2.398&	0.114&	0.124&	30.715&	\textless0.001&	0.644&	4.643&	0.008&	0.215\\
		& Average Speed&	0.821&	0.416&	0.046&	0.200&	0.761&	0.012&	9.075&	\textless0.001&	0.348\\
		& Std. Speed&	0.304&	0.68&	0.018&	0.904&	0.404&	0.050&	1.366&	0.266&	0.074\\
		
		\bottomrule
	\end{tabular}
	
	\label{table 2}
\end{table*}

\section{Result}
Driving performance data were analyzed using repeated measures ANOVA with input modality (baseline, manual, and speech) and output modality (combined, text, and audio) as within-subject variables. The results were corrected for sphericity violations where necessary by use of the Greenhouse–Geisser modification \cite{greenhouse_methods_1959}. Where significant main effect or interaction appeared, post hoc tests following the conservative bonferonni correction was further performed for the pair-wise comparisons. A significance level of .05 was adopted for the significance tests. The ANOVA results are shown in Table 2.

\subsection{Hazard Events}
Input modality and its interaction with output modality had a statistically significant effect on BrakeRT (Table 2). The analyses of simple effects showed that with combined output, manual input resulted in slower reaction compared to baseline (p\textless0.001); with visual output, manual input led to longer reaction time compared to both speech input (p=0.009) and baseline (p=0.043); but with audio output no significant effect was found. Additionally, when manual input was adopted, combined and visual output both produced longer BrakeRT compared to audio output (combined: p=0.023; visual: p=0.035); however, no such effect was found when speech input was adopted. These results indicate manual input severely degrades hazard detection when textual order information is presented, which is further validated by the fact that the longest reaction time appeared with the visual-manual combination (M = 2.945, SD = 0.130).

ANOVA results indicated that output modality and its interaction with input modality both had a statistically significant effect on the approaching speed (Table 2). The analyses of simple effects revealed that with speech input, audio output produced higher speed compared to combined input (p=0.043); but no such effect was found under other input modalities. The highest speed occurred with the audio-speech combination (M=62.365, SD=1.512). These results could be interpreted as the drivers' adaption behavior towards possible hazard events.

Results of ANOVA indicated that only the interaction between input modality and output modality imposed a significantly effect on MinHWT (Table 2). The analyses of simple effects revealed that manual input resulted in shorter MinHWT compared to speech input (p=0.027) and baseline (p\textless0.001) with combined output; but such effect did not appear under other output modalities. The shortest MinHWT (M=0.721, SD=0.095) occurred under combined output and manual input. These results manifest that combined output exacerbate negative influence caused by manual input upon the risk of accident. Results are shown in Figure 4.

\begin{figure*}
	\centering
	\setlength\tabcolsep{1.5pt} 
	\begin{tabular}{cc} 
		\includegraphics[width=0.5\textwidth]{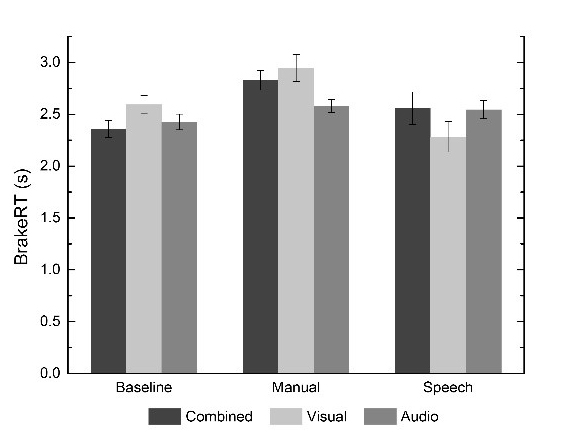} &
		\includegraphics[width=0.5\textwidth]{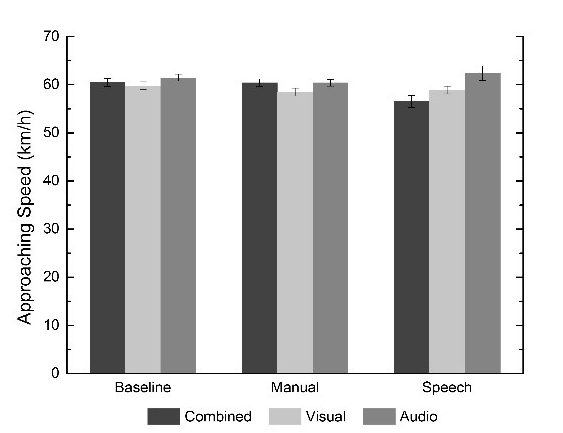} \\
		(a) brake reaction time & (b) Approaching speed \\
		\multicolumn{2}{c}{\includegraphics[width=0.5\textwidth]{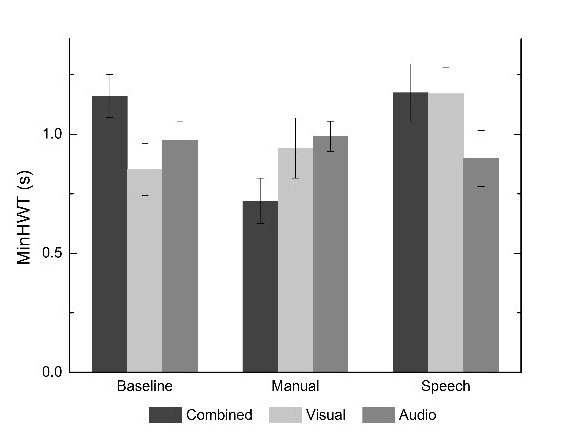}}\\
		\multicolumn{2}{c}{(c) Minimum headway time} \\
	\end{tabular}
	\caption{Driving performance in hazard events}
\end{figure*}


\begin{figure*}
	\centering
	\setlength\tabcolsep{1.5pt} 
	\begin{tabular}{cc} 
		\includegraphics[width=0.5\textwidth]{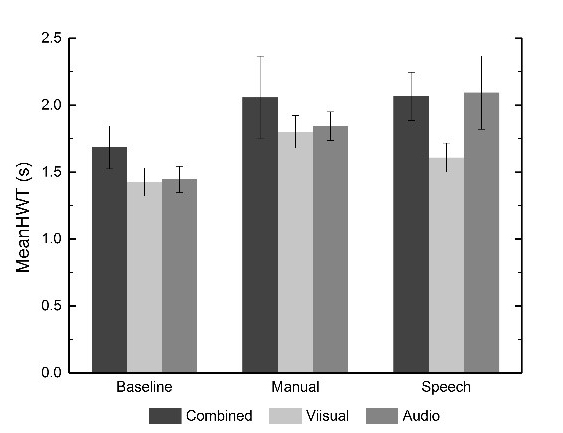} &
		\includegraphics[width=0.5\textwidth]{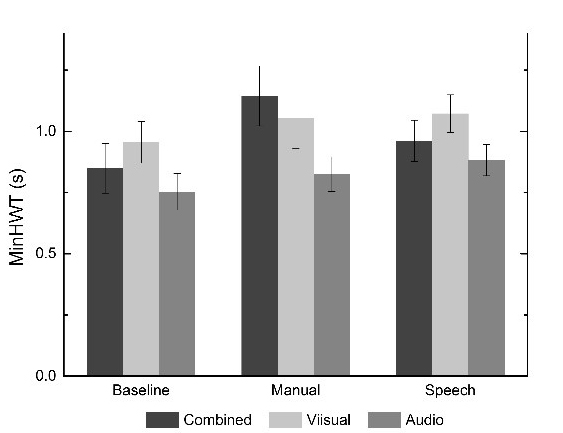} \\
		(a) Mean headway time & (b) Minimum headway time \\
		\multicolumn{2}{c}{\includegraphics[width=0.5\textwidth]{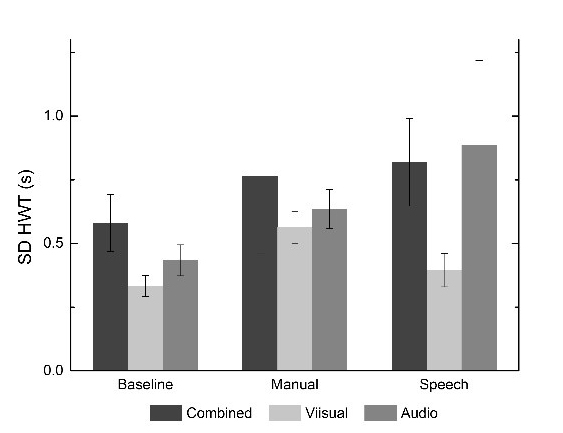}}\\
		\multicolumn{2}{c}{(c) Standard deviation of headway time} \\
	\end{tabular}
	\caption{Driving performance in hazard events}
\end{figure*}


\begin{figure*}
	\centering
	\setlength\tabcolsep{1.5pt} 
	\begin{tabular}{cc} 
		\includegraphics[width=0.5\textwidth]{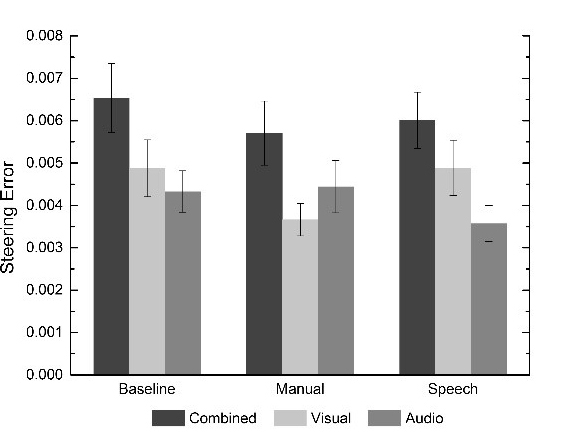} &
		\includegraphics[width=0.5\textwidth]{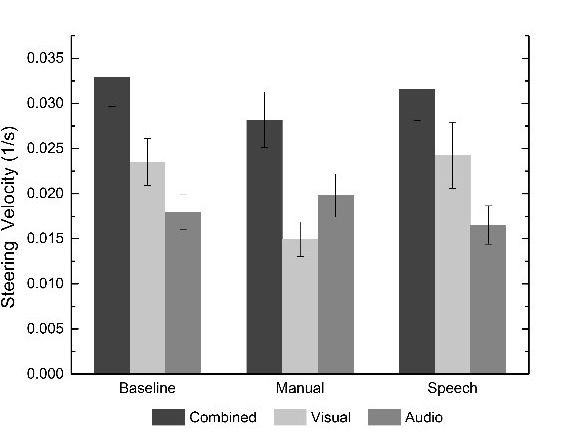} \\
		(a) Steering error & (b) Steering velocity \\
			\includegraphics[width=0.5\textwidth]{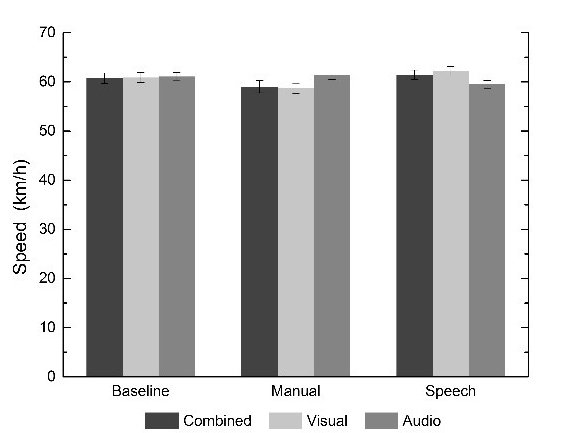} &
		\includegraphics[width=0.5\textwidth]{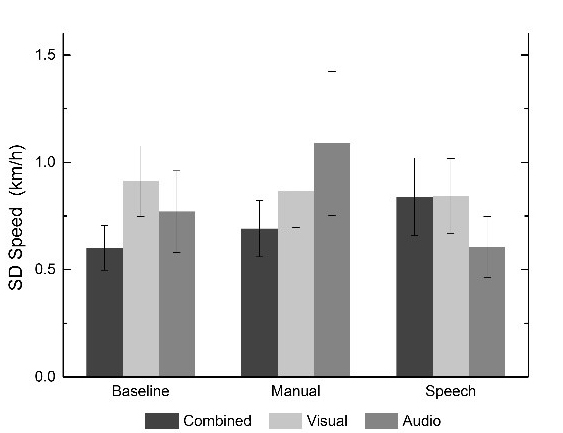} \\
		(c) Speed & (d) Standard deviation of speed \\
	\end{tabular}
	\caption{Driving performance in lane changing task}
\end{figure*}

\subsection{Car Following Task}
The ANOVA results indicated that input modality had a significant effect on the mean headway time (Table 2). Manual and speech input both led to significant increases in drivers' MeanHWT compared to baseline (manual: p=0.006; speech: p=0.046). In the baseline drive, drivers strictly maintained the 1.5s gap from the leading vehicle (M=1.519, SD=0.098). While in the manual and speech drives, the MeanHWT increased by about 0.4s (manual: M=1.901, SD=0.141; speech: M=1.923, SD=0.122). These results could be interpreted as that drivers tend to maintain a longer headway time to the leading vehicle when distracted by secondary tasks.

For the minimum headway time, only output modality showed a significant effect (Table 2). Drivers produced shorter MinHWT when audio output was presented compared to with combined output (p=0.013) and visual output (p=0.002). These results reveal that audio output leads to more aggressive driving style compared to other output modalities. However, ANOVA failed to find any significant effect on HWT variability. Results are shown in Figure 5.

\subsection{Lane Changing Task}
The ANOVA results indicated that output modality and its interaction with input modality respectively had a significant and marginally significant effect on steering error (Table 2). Combined output led to significant increases in drivers' steering error compared to visual output (p=0.001) and audio output (p\textless0.001). With respect to the steering velocity, output modality and its interaction with input modality both showed a statistically significant effect (Table 2). Analyses of simple effects revealed that regardless of the different input modalities, drivers produced quicker steering when combined output were present compared to visual and audio output. These results reveal that combined output modality makes drivers' steering control more abrupt and quicker. Furthermore, with visual output modality, manual input led to slower steering compared to speech input (p=0.010) and baseline (p=0.001); however, with other output modalities, no significant effect was found. The slowest steering velocity appeared with the visual-manual combination; while the smoothest steering occurred with the audio-speech combination.

For the average speed, only the interaction between input modality and output modality showed a statistically significant effect (Table 2). Analyses of simple effect showed that with manual input, audio output produced higher speed compared to combined output (p=0.003) and visual output (p=0.006); with speech input, audio output in return resulted in lower speed compared to combined output (p=0.026) and visual output (p=0.035); while no significant effect was identified under baseline condition. Additionally, with visual output, manual input led to lower speed compared to speech input (p=0.023). The highest (M=62.201, SD=0.987) and lowest speed (M=58.704, SD=1.032) occurred under visual-speech and visual-manual modality respectively. These results could also be interpreted as the drivers' adaptive behavior towards the distraction related to secondary task. Results are shown in Figure 6.

In the experiment, drivers missed a total of 19 lane change signs, 10 of which were in the task session, the other 9 were in the ordinary driving session. An additional chi-square goodness-of-fit analysis of the number of missed lane change signs on different input modalities found that manual input led to more, although not significant, missed lane change signs compared to other input modalities  ($\chi^2$(1, n = 19) =3.800, p=0.150). In addition, under condition of visual output, only manual input produced 4 missed lane changes, which was the largest in all of the input and output combinations, as shown in Table 3.

\begin{table}[!htb]
	\small\sf\centering
	\caption{Distribution of missed lane-changing signs during task sessions}
	\begin{tabular}{lllll}
		\toprule
		&	 Combined & Visual & Audio & Total \\
		\midrule
		Baseline & 3 & 0 & 0 & 3 \\
		Manual & 2 & 4 & 0 & 6 \\
		Speech & 0 & 0 & 1 & 1 \\
		Total & 5 & 4 & 1 & 10 \\
		\bottomrule
	\end{tabular}\\
	\label{table 3}
\end{table}

\section{Discussion}
Drivers maintained a shorter minimum headway time (M=0.821, SD=0.055) under the condition of audio output modality compared to combined and visual modalities in car following task. This result could be interpreted as that when visual resources were occupied by the secondary tasks, drivers tend to reduce the resources demand of the driving task by adopting conservative driving maneuver; however, with audio output, drivers are confident in handling the driving and secondary tasks simultaneously so that adopt an aggressive one. These results accord with the previous findings that visual distraction imposed a larger effect on drivers compared to auditory distraction, and thus led to more adaptive behaviors \cite{jensen_studying_2010,moldenhauer_effect_2003}.

Compared to baseline drive, drivers maintained longer mean time headway in the car following task in manual and speech drives. Such difference should be ascribed to the integrative effects of both the input and output factors considering that drivers were not only free of operating the application but also not disturbed by the order information in the baseline drive. Drivers might realize the increased difficulty to observe and follow every accelerating and decelerating action of the leading vehicle because of the presence of secondary task, and so adopt a cautious strategy – keeping a larger safety margin in case of overlooking the decelerating action and resulting in a rear end collision. Similar compensative behaviors of drivers under increased workload were also reported in a number of studies \cite{liang_combining_2010,ranney_effects_2005,harbluk_using_2007,tsimhoni_address_2004}.

In current taxi-hailing applications, the same order information was presented to the drivers in the combined visual-audio output modality with the purpose of letting drivers to choose a relatively less driving-interfering way, in most case the audio modality, in certain driving contexts. However, the experimental results run contrary to such expectation. The effects of combined output modality differed from those of audio output modality while displayed a similar pattern with those of visual output modality in all dependent variables. This finding accords with the previous studies on other IVIS in that drivers inclined to pay attention to the visual output although the less detrimental modality – audio output – was provided (Green et al., 1993; Jensen et al., 2010). However, previous studies failed to reach a consensus on the effects of audio output when attached to the visual output (Jensen et al., 2010; B. Lee et al., 2014; Richard et al., 2002). In this study, combined and visual output modality placed comparable effects, which significantly differed from those imposed by audio output modality, on drivers' longitudinal control (MinHWT in car following task and speed in lane changing task with manual input modality) and hazard detection (brake reaction time in hazard event and MinHWT in the hazard event with manual input modality). These results suggest that under the combined output modality, visual distraction dominates the effects on longitudinal control and hazard detection decrements; while the role played by the auditory distraction is approximately negligible. However, the auditory distraction exacerbates the negative effects of visual distraction on the lateral control, reflected in that combined output modality produced significantly quicker and more abrupt steering compared to both visual and audio modalities.

According to the hypothesis, the simultaneous occupation of visual and manual resources most severely impairs the driving performance, which is supported by the results of the brake reaction time and minimum headway time in hazard events. Besides, the speed in lane changing task suggests that visual-manual modality also causes the most adaptive behaviors. One explanation concerns with the multiple resource theory, which defines attentional resources in four dimensions: processing stage (i.e., perception/central processing vs. response), sensory modalities (i.e., visual or auditory), processing code (i.e., the process of analogue/spatial vs. categorical/verbal information, and channels of visual information (i.e., focal vs. ambient vision) \cite{horrey_driving_2004,wickens_multiple_2002}. Driving tasks makes major demands on the resources associated with visual perception, spatially coded-working memory, and motor response, while minor demands on auditory perception and verbally coded-working memory, or verbal response \cite{liang_combining_2010}. Handling a taxi-hailing in visual-manual modality competes for the same resources with the driving tasks so that degrades the driving performance; drivers are aware of such competition and then modulate the resources distribution by reducing their speed. Somewhat surprisingly the lateral control was not most seriously impaired by the visual-manual modality, and the steering velocity even reached the least value under this modality. One possible explanation is that drivers missed the most lane changing signs under visual-manual modality, which means less steering actions are taken and thus leads to a lower average steering velocity.

In conformity with previous studies, speech input modality is overall less distracting to drivers than manual input modality \cite{maciej_comparison_2009,tsimhoni_destination_2002,tsimhoni_address_2004}. Moreover, according to the MRT the combination of audio output and speech input is expected to be the least detrimental to driving performance, because auditory perception and verbal response are barely needed by the driving task. However, the optimum combination of input and output modality varies in terms of different dependent variables. For lateral control, audio-speech modality does produce the least steering error. With regard to longitudinal control, audio-speech modality results in highest approaching speed in hazard event; while in lane changing task, the highest average approaching speed occurs under visual-speech modality. This difference may be caused by the distinction between the driving contexts of hazard events and lane changing tasks \cite{tivesten_driving_2014}. With respect to the hazard detection, the drivers fastest react to the hazard events under speech-visual modality, and maintain the longest minimum headway time under speech-combined modality, which could be regarded as speech-visual modality because their minor difference (speech-combined: M=1.175, SD= 0.119; speech-visual: M=1.171, SD= 0.108) and that visual component dominates the combined modality as stated above. The reason might associate with the drivers' gaze concentration. Unlike the lane changing task and car following task, the hazard events occur during normal driving, which does not require drivers' visual attention to any specific spot of the driving scene. So drivers tend to concentrate their gaze in the center of the driving scene when order of audio-speech modality introduces cognitive distraction to them, and thus their ability to detect targets across the entire driving scene is diminished \cite{recarte_mental_2003,victor_keeping_2005}. When handling order of visual-speech modality, drivers incline to consciously saccade the whole driving scene after short off-road glance (to get the order information) in order to comprehend the new driving contexts, and thus observe the pulling-out car sooner.

The study conducted in driving simulators has the advantage of being highly controllable and repeatable. However, the generality of the conclusions drew from simulator-based studies is limited by the extent to which the simulator presents the drivers with actual driving situations. Another limitation of this study is the relatively small sample size, which undermines the statistical power. The participants in this study were young drivers, and the sample can be extended to include mature drivers who experienced in both driving and handling taxi-hailing orders. Despite of such limitations, the results generalized from this study are useful for understanding the mechanism that how the interaction with taxi-hailing application influence the driving performance. It helps in improving the design of not only the taxi-hailing application itself, but also other input-output balanced IVIS.


\small \bibliography{IEEEabrv,app_reference_arxiv}

\end{document}